\documentclass[aps,prl, 10pt, twocolumn, superscriptaddress,nofootinbib]{revtex4}
\usepackage{subeqn}

\newcommand{\be}{\begin{equation}}
\newcommand{\ee}{\end{equation}}
\newcommand{\bea}{\begin{eqnarray}}
\newcommand{\eea}{\end{eqnarray}}

\begin{document}

\title{Comment on ``Model for Gravity at Large Distances"}

\author{Tongu\c{c} Rador}\email[]{tonguc.rador@boun.edu.tr} 
\affiliation{\.Izmir Institute of Technology, Department of Physics \\ Urla 35430, \.Izmir, Turkey}
\affiliation{Bo\u{g}azi\c{c}i University, Department of Physics \\ Bebek 34342, \.Istanbul, Turkey}

\author{ Sava\c{s} Arapo\u{g}lu } \email[]{ arapoglu@itu.edu.tr }
\affiliation{\.Istanbul Techincal University, Physics Engineering Department\\ Maslak 34469, \.Istanbul, Turkey}

\author{\.Ibrahim Semiz}\email[]{semizibr@boun.edu.tr}
\affiliation{Bo\u{g}azi\c{c}i University, Department of Physics \\ Bebek 34342, \.Istanbul, Turkey}

\begin{abstract}
We correct a sign mistake in the work mentioned in the
title; explore consequences on energy conditions in the relevant
context, and make a suggestion on the introduced parameter.
\end{abstract}

\pacs{}

\maketitle

% \section{}

Recently Grumiller \cite{Grumiller}, starting from a simple set of assumptions, proposed the metric\footnote{We take $\Lambda=0$ without losing generality of our arguments.}

\begin{subequations}{\label{one}}
\bea
ds^{2} &=& -K^{2}dt^{2} + \frac{dr^{2}}{K^{2}}+r^{2}\left(d\theta^{2}+\sin^{2}\theta d\phi^{2}\right),\\
K^{2} &=& 1-2 \frac{MG}{r}+ 2 b r,
\eea
\end{subequations}

\noindent as a somewhat general framework to approach various systems with anomalous accelerations such as the rotation curves of spiral galaxies and the Pioneer anomaly. 
It is stated that $b$ comes in as an arbitrary constant depending on the system under study and that for $b>0$ and of the order of inverse Hubble length, a qualitative understanding of the mentioned anomalies are possible. It is also stated that the effective energy-momentum tensor resulting from Eqs.(\ref{one}) is that of an anisotropic fluid obeying the equation of state $p_{r}=-\rho$ and $p_{\theta}=p_{\phi}=p_{r}/2$ with
\be{\label{onunki}}
\rho=\frac{4 b}{\kappa r}\;,
\ee
where $\kappa$ is the (positive) gravitational coupling constant, i.e. the constant in the Einstein equation $G_{\mu\nu} =\kappa T_{\mu\nu}$.

While we agree on the equation of state we disagree on the sign of $\rho$; the metric in Eqs.(\ref{one}) yields\footnote{We use MTW \cite{MTW} sign conventions, but of course the signs of $\rho$ and $p$'s are the same in all commonly used conventions.}
\be{\label{bizimki}}
\rho=-\frac{4 b}{\kappa r}\;.
\ee
The effective potential formalism for the geodesic equation of a test particle is given by
\be
\left(\frac{dr}{d\lambda}\right)^{2}+2V_{\rm eff}(r)={E^{2}-\epsilon},
\ee
with
\be{\label{veff}}
V_{\rm eff}(r)=-\epsilon\frac{MG}{r}+\frac{L^{2}}{2r^{2}}-MG\frac{L^{2}}{r^{3}}+\epsilon br+b\frac{L^{2}}{r},
\ee
where units are chosen such that $c=1$, $\lambda$ is the affine parameter along the geodesic and the constants of motion are $E=K^{2}dt/d\lambda$ and $L=r^{2}d\phi/d\lambda$. Also, for massive and massless particles we have $\epsilon=1$ and $\epsilon=0$ respectively.

It is still true that for $b>0$ the effect of $br$ is a constant anomalous acceleration towards the center for objects moving non-relativistically , {\em despite the negative energy density}. This follows,
because the effective potential is derived from the metric directly;
and can also be seen from the Raychaudhuri equation specialized to a
collection of test particles initially at rest in a small volume of
space \cite{Carroll}:
\be\label{raycha1}
\frac{d\theta}{d\tau}=-4\pi G(\rho+p_{r}+p_{\theta}+p_{\phi})=4\pi G\rho\;,
\ee
where $\theta$ is the quantity called {\em expansion} and $\tau$ is the proper-time along geodesics followed by the test particles\footnote{The LHS of Eq.\ref{raycha1} can also be written as $\ddot{V}/V$ where $V$ denotes the volume occupied by the test particles. See \cite{baezbunn} for a nice introduction to general relativity where Raychaudhuri's equation is placed at the center of exposition.}. The second equality follows from the peculiar equation of state described in \cite{Grumiller} and confirmed here. The negativity of the derivative of expansion along geodesics shows that gravity is attractive for a given fluid; this is the case here because of Eq.(\ref{bizimki}), for positive $b$.

The negative energy density naturally leads us to question if the fluid violates any of the so-called {\em energy conditions}\footnote{We use the definitions in \cite{Carroll} for the energy conditions.}, the compatibility with which is generally taken as a measure of physical reasonableness. The weak energy condition (WEC) requires $\rho \geq 0$ and $\rho+p_{i} \geq 0$; the strong energy condition (SEC), $\rho+\sum{p_{i}} \geq 0$ and $\rho+p_{i} \geq 0$; and the dominant energy condition (DEC), $\rho \geq |p_{i}|$; it is easily seen that our fluid violates all three\footnote{One can easily  find that the null energy condition (NEC) and the null dominant energy condition (NDEC) are also violated.}.

On one hand, one might say that the violation of energy conditions means that the model is not very physically reasonable; but on the other hand, we are talking about an {\em effective} fluid, not necessarily a real one. Also, the attractive nature of the fluid (as confirmed by application of the Raychaudhuri's equation) in the face of these violations serves as an example of a delicate fact about SEC: while SEC ensures that gravity is attractive it does not encompass all {\em attractive} gravities.

Finally, we would like to point out a possibility for the relation between $b$ and the system under consideration: The fluid is attractive; in fact, the $1/r$ dependence of the density and pressures show that it clusters around the central mass. Though speculative at this point, it seems reasonable that bigger masses will accumulate more fluid, i.e., $b$ will be a monotonically increasing function of $M$. On the other hand the very meaning of $M$ next to $b$ is questionable
because one has to match the metric in Eq.(\ref{one}) to the metric of the interior system (star or galaxy), the matching conditions will undoubtedly yield a relation between the integral of the energy density inside the interior system which we may call $M_{s}$ and the parameters of the metric outside; $M$ and $b$. We leave the quantitative
analysis of these points for  future work.


\begin{thebibliography}{100}
\bibitem{Grumiller} D. Grumiller, ``Theory of Gravity at Large Distances", Phys. Rev. Lett. 105, 211303 (2010), arXiv:1011.3625.
\bibitem{Carroll} S. Carroll, ``Spacetime and Geometry", Addison-Wesley, ISBN 0-8053-8732-3.
\bibitem{baezbunn} J.C. Baez and E.F. Bunn, ``The Meaning of Einstein's Equation",       Amer. Jour. Phys. 73 (2005), 644-652, arXiv:gr-qc/0103044.
\bibitem{MTW} C.W. Misner, K.S. Thorne and J.A. Wheeler, ``Gravitation", Freeman, ISBN 0-7167-0344-0 (pbk).

\end{thebibliography}
\end{document}